# Reverse Doping Asymmetry in Semiconductor Thin Films Using External Voltage

Kai Liu[1], Zhibin Yi[1], and Guangfu Luo[1,2,*]

[1]Department of Materials Science and Engineering, Southern University of Science and Technology, Shenzhen, Guangdong 518055, China

[2]Guangdong Provincial Key Laboratory of Computational Science and Material Design, Southern University of Science and Technology, Shenzhen, Guangdong 518055, China.

*E-mail: luogf@sustech.edu.cn

## Abstract

Doping asymmetry is a notable phenomenon with semiconductors and a particularly longstanding challenge limiting the applications of most wide-band-gap semiconductors, which are inherent of spontaneous heavy n- or p-type doping because of their extreme band edges. This study theoretically shows that by applying a proper external voltage on materials during their growth or doping processes, we can largely tune the band edges and consequently reverse the doping asymmetry in semiconductor thin films. We take zinc oxide as a touchstone and computationally demonstrate that this voltage-assisted-doping approach efficiently suppresses the spontaneous n-type defects by around four orders under three distinct growth conditions and successfully generates p-type zinc oxide up to the lowest acceptor levels. The proposed approach is insensitive to materials, growth conditions, or defects origins, and thus offers a general solution to the doping asymmetry in semiconductor thin films.

# I. Introduction

Wide-band-gap (WBG) semiconductors possess a broad range of fascinating applications [1-3], such as blue lasers, high-power/high-temperature/high-frequency devices, transparent electrodes, and potentially room-temperature spintronics. However, decades of studies have clearly demonstrated that most WBG semiconductors are inherent in forming heavy n- or p-type doping and meantime suffer from overwhelming difficulties to achieve the opposite doping type [4-6]. This phenomenon is known as the doping asymmetry, which renders an ideal design based on the homo-junctions of WBG semiconductors unreachable and is particularly unfortunate for those with wafer-size single crystals, such as ZnO and $Ga_2O_3$. The doping asymmetry originates from the fact that a wide band gap usually means either a very high conduction band maximum (CBM), e.g. -0.2 eV for diamond [7,8], or a very low valance band maximum (VBM), e.g. -7.8 eV for ZnO and -8.9 eV for $Ga_2O_3$ [9,10], which renders the formation of n-type or p-type defects thermodynamically unfavorable and the corresponding ionization energy levels relatively deep in the band gap. Such feature is clearly shown in Fig. 1, which compares band edges among the typical spontaneous n- or p-type WBG semiconductors and conventional semiconductors [11,12]. The only two commercialized WBG semiconductors, GaN and SiC, fortunately possess band edges close to those of GaAs and Si. These extreme band edges are intrinsic with WBG semiconductors and thus explain their long-standing doping asymmetry.

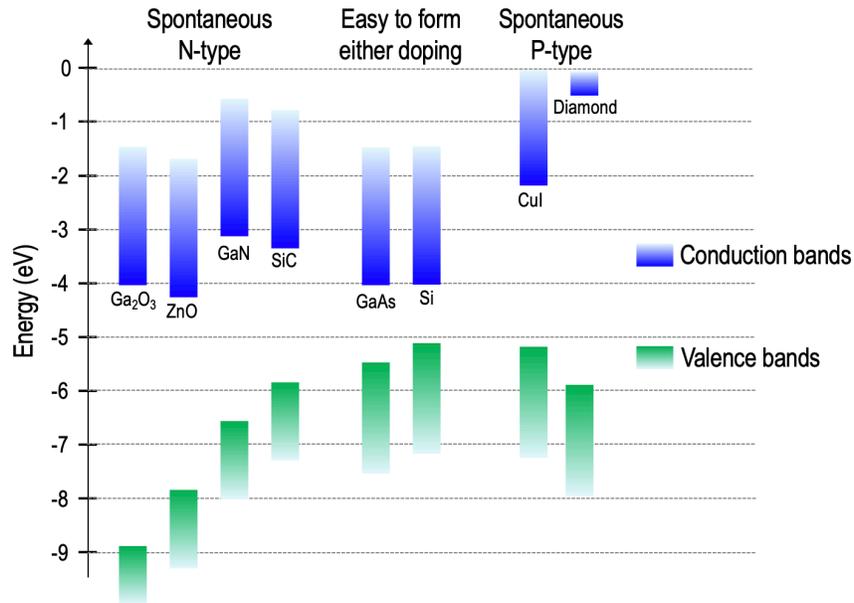

FIG. 1. Band edges of typical WBG semiconductors and those of conventional semiconductors relative to the vacuum level. Original data were obtained from literature [7,8].

According to the defect thermodynamics, the concentrations of charged defects in crystals are governed by the Fermi energy and the chemical potentials under specific growth conditions [13,14]. The two factors are typically coupled through the charge neutrality condition and the defect concentrations can be determined solely by the growth conditions. Because of the narrow optimal growth windows and the compensation effects of spontaneous defects [15-17], traditional attempts to achieve the nonspontaneous doping type by changing growth conditions have achieved very limited success [18]. In this study, we propose to tune the defect thermodynamics by relatively shifting the band edges of semiconductors using external voltages during the growth or doping processes. We derive the theoretical formula of this approach, propose a schematic device, and choose ZnO as a classical example to demonstrate this approach using first-principles calculations. Our results reveal distinct defect origins of the spontaneous n-type doping for three representative growth methods of ZnO and the proposed voltage-assisted-doping approach successfully generates p-type doping up to the lowest acceptor level.

## II. METHODS

All the first-principles calculations in this work were based on the density functional theory (DFT), as implemented in the Vienna Ab-initio Simulation Package [19]. We adopted the exchange-correlation functional of HSE06 [20] as well as the Hubbard model with an effective U-J value of 6 eV for the 3d electrons in Zn. This approach gives an overall excellent description of the lattice parameters, band gap, and formation energy of ZnO bulk [21]. By contrast, the typical method of using HSE functional with 36% of Hartree-Fock exact exchange overestimates the formation energy of ZnO bulk by 22% [21]. To describe the special relativity effects in Pb, the spin-orbit coupling is included for the relevant calculations. A plane-wave energy cutoff of 435 eV was used, together with following projector augmented wave potentials: Zn_GW ($3d^{10}4p^2$) for Zn, O_s_GW ($2s^22p^4$) for O, H_GW ($1s^1$) for H, N_GW ($2s^22p^3$) for N, P_GW($3s^23p^3$) for P, As_GW($4s^24p^3$) for As, Ag_GW($5s^14d^{10}$) for Ag, and Pb_sv_GW ($5s^26s^25p^66p^25d^{10}$) for Pb. The Brillouin zone was sampled with a 2 × 2 × 2 gamma-centered mesh.

A 72-atom supercell of wurtzite ZnO is used to simulate thirty eight defects in total, including seven native point defects: oxygen and zinc vacancies ($V_O$ and $V_{Zn}$), interstitials ($O_{i1}$, $O_{i2}$, and $Zn_i$), and antisites ($O_{Zn}$ and $Zn_O$); eight native pair-defects: $V_O$+$V_O$, $V_{Zn}$+$V_{Zn}$, $V_O$+$O_i$, $V_{Zn}$+$Zn_i$, $V_O$+$V_{Zn}$, $V_O$+$O_{Zn}$, $V_O$+$Zn_i$, and $V_O$+$Zn_O$; seven H-related defects: $H_{Zn}$, $H_O$, $H_i$, $V_{Zn}$+2H, $H_i$+$O_{Zn}$, $H_i$+$V_{Zn}$, and $H_i$+$O_i$; two defects related to the impurity lead: $Pb_{Zn}$ and $Pb_{Zn}$+$V_O$; and fourteen defects related to the typical p-type dopants: $N_i$, $N_{Zn}$, $N_O$, $N_{Zn}$+$V_{Zn}$, $N_{Zn}$+$2V_{Zn}$, $P_O$, $P_{Zn}$, $As_O$, $As_{Zn}$, $As_{Zn}$+$V_{Zn}$, $As_{Zn}+2V_{Zn}$, $Ag_i$, $Ag_{Zn}$, and $Ag_{Zn}+V_O$. These defects are chosen with consideration of the known defect types, the dominant impurities in the examined growth methods, and the typical dopants. Symmetry is broken to allow a possible Jahn–Teller distortion around the defects. The *ab initio* method proposed by Freysoldt, Neugebauer and Van de Walle (FNV correction) is adopted to reduce the image charge interaction and adjust the potential alignment between the perfect and defected structures [22,23]. Our tests indicate that the error in defect formation energy after the FNV correction is limited to ~0.1 eV [21]. Table 1 summarizes the typical growth conditions and the corresponding chemical potentials for the MBE growth [24], hydrothermal growth [18,25,26], and pressurized melt growth [27,28] of ZnO.

**Table 1.** Experimental growth temperature, $T_G$, partial pressure of oxygen, $P_{O2}$, partial pressure of water vapor, $P_{H2O}$, and concentration of lead impurity for the MBE growth, hydrothermal growth (HTG), and pressurized melt growth (PMG) of ZnO, and the corresponding chemical potentials used in this work. The chemical potentials for the Zn-rich and O-rich conditions are given for comparison; all growth conditions are organized in decreasing Zn chemical potential.

|  | Zn-rich[†] | MBE[‡] | HTG[$] | PMG[§] | O-rich[†] |
|---|---|---|---|---|---|
| $T_G$ (K) | — | 692 | ~650 | 2200 | — |
| $P_{O2}$ (atm) | — | $1.3\times10^{-7}$ | $~5\times10^{-31}$ | ~50 | — |
| $P_{Zn}$ (atm) | — | $2\times10^{-4}$ | — | — | — |
| $P_{H2O}$ (atm) | — | — | ~1085 | — | — |
| Pb (cm$^{-3}$) | — | — | — | $~3.6\times10^{17}$ | — |
| N (cm$^{-3}$) |  | $~10^{12}$ | $~10^{12}$ | $~10^{12}$ |  |
| $\mu_{Zn}$ (eV) | -0.70 | -1.10 | -1.98 | -3.51 | -4.16 |
| $\mu_O$ (eV) | -10.60 | -8.03 | -9.46 | -9.18 | -7.14 |
| $\mu_H$ (eV) | — | — | -4.09 | — | — |
| $\mu_N$ (eV)[∥] | — | -5.48 | -4.32 | -8.97 | — |
| $\mu_{Pb}$ (eV)[∥] | — | — | — | -7.87 | — |

[†]$\mu_{Zn}$ under Zn-rich condition equals the total energy per atom of Zn metal and $\mu_O$ under the O-rich condition corresponds to the total energy per atom of $O_2$ molecule; the equilibrium condition of $\mu_O + \mu_{Zn} = \mu_{ZnO}$ is used to obtain the other chemical potential under the same condition, where $\mu_{ZnO}$ is the average energy per formula of ZnO bulk.

‡$\mu_{Zn}$ and $\mu_O$ are the Gibbs free energy per atom of the Zn and $O_2$ beams, respectively.

$\mu_H$ and $\mu_O$, respectively, correspond to the free energy per atom of $H_2$ and $O_2$ under the equilibrium conditions of water decomposition at 650 K and 1085 atm. $\mu_{Zn}$ is calculated using the equilibrium condition of $\mu_O + \mu_{Zn} = \mu_{ZnO}$ at 650 K.

§$\mu_O$ is the Gibbs free energy of $O_2$ per atom under the experimental conditions and $\mu_{Zn}$ is calculated using the equilibrium condition of $\mu_O + \mu_{Zn} = \mu_{ZnO}$ at 2200 K.

∥$\mu_{Pb}$ and $\mu_N$ are evaluated self-consistently, so the total content of N and Pb equal ~$10^{12}$ cm$^{-3}$ and $3.6 \times 10^{17}$ cm$^{-3}$, respectively.

## III. Results and Discussions

In this study, we first lay a theoretical foundation of tuning the defect thermodynamics using external voltage. After that, we computationally examine the point defects in ZnO bulk grown by MBE, HTG, or PMG, and reveal the defect origins of the spontaneous n-type doping. Then, we apply our voltage-assisted-doping approach to the different growth methods of ZnO bulk and demonstrate its effectiveness of reversing the doping asymmetry. Finally, we propose an experimental device to realize the voltage-assisted-doping approach.

### A. Theoretical Model of the Voltage-Assisted-Doping Approach

Our start point is the classical definition of formation energy for a dilute point defect $A_B^q$, as illustrated in Eqn. 1 [13,14].

$$E_f(E_F) = E_{tot}(A_B^q) - E_{tot}(perfect) + \mu_B - \mu_A + q(E_{VBM} + E_F) \quad (1)$$

where $A_B^q$ refers to a defect with element $A$ occupying a lattice site $B$ and possessing a charge state $q$; $E_{tot}(A_B^q)$ and $E_{tot}(perfect)$ are the total energies of systems with the charged point defect and the perfect crystal, respectively. $\mu_A$ and $\mu_B$ are the chemical potentials of element $A$ and $B$ under experimental conditions, respectively. $E_{VBM}$ is the VBM energy in perfect crystal, and $E_F$ is the Fermi level relative to $E_{VBM}$. In a supercell calculation, an extra term to correct the finite-size errors for charged defects [22,23] should be added to $E_{tot}(A_B^q)$. The last term of Eqn. 1 explains why the extreme band edges of WBG semiconductors (Fig. 1) can easily lead to the doping asymmetry.

In this study, we propose to tune the band edges of semiconductors using an external voltage. Figure 2a schematically shows the critical energy levels of a n-type semiconductor thin film under a positive external voltage for the model in the inset. The spatial distributions of electrostatic potential, Fermi level, and band edges in the model can be solved using the Gaussian's law, with consideration of the dopant concentrations and free charge carriers [29]. The ground terminal is subject to the Dirichlet boundary condition, while the insulator/semiconductor interface adheres to the continuity of electric displacement field. Numerical simulations with the COMSOL package [30] indicate that excess electrons can accumulate in the high-potential region and induce a downward band bending, namely $U_G \equiv E_{CBM}(V_{ext}) - E_{CBM}(0)$. Detailed examination of n-type ZnO films with different thickness and donor concentrations reveals that, under the high growth temperatures of ZnO, $U_G$ can be tuned till the Fermi level touches the CBM of ZnO, and the largest magnitude is about 1.5 eV (Fig. 2b) or nearly half of the band gap [21].

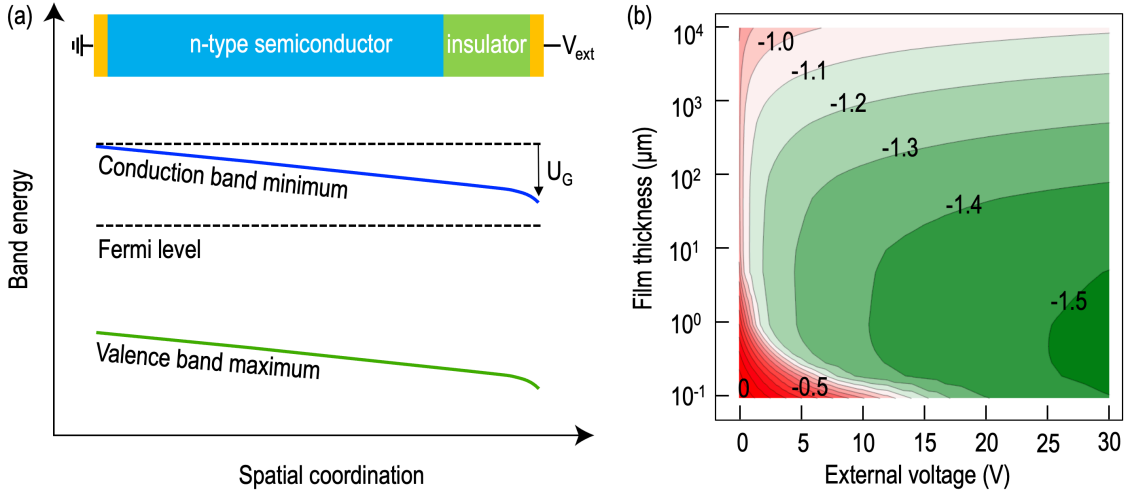

FIG. 2. (a) Schematic spatial dependence of band edges in a n-type semiconductor thin film under a positive external voltage. Inset shows a schematic device. (b) External voltage-induced band bending for n-type ZnO films with a donor concentration of $10^{14}$ m$^{-3}$ at 650 K, which is a lower border of the growth temperatures of ZnO. Labels around the contours are in unit of eV.

With such band bending, charged defects will gain new equilibrium distributions. For instance, the downward band bending in Fig. 2a will attract/generate negatively-charged or p-type defects but repel/depress the positively-charged or n-type defects. Quantitatively, the electrostatic energy of defect $A_B^q$ changes by $-qU_G$, where the minus sign is added because $U_G$ is referenced to electrons. The energies of perfect bulk, chemical potentials, and Fermi level, however, remain unchanged. Therefore, the total change of defect formation energy under the band bending equals $-qU_G$ and the defect formation energy is updated from Eqn. 1 to Eqn. 2.

$$E_f(E_{FG}, U_G) = E_{tot}(A_B^q) - E_{tot}(perfect) + \mu_B - \mu_A + q(E_{VBM} + E_{FG} - U_G), \qquad (2)$$

where $E_{FG}$ is the Fermi level under the gated voltage. Since Eqn. 2 is mathematically equivalent to shifting the $E_{VBM}$ in Eqn. 1 by $-U_G$, the defect formation energy curves based on Eqn. 2 can be obtained by simply shifting the $E_{VBM}$ for the curves based on Eqn. 1.

With the defect formation energy, the corresponding defect concentration can be calculated through Eqn. 3.

$$c(T_G, E_{\text{FG}}, U_G) = n_0 e^{-E_f(E_{\text{FG}}, U_G)/(k_B T_G)}, \qquad (3)$$

where $n_0$ is the number density of potential defect sites and $k_B$ the Boltzmann constant. The only variable in Eqns. 2 and 3 that cannot be easily obtained is $E_{FG}$, which, however, can be solved self-consistently according to the neutrality requirement, namely that the charges of all defects and free charge carriers add up to zero, as shown in Eqn. 4.

$$n_h(T_G, E_{\text{FG}}) - n_e(T_G, E_{\text{FG}}) + \sum_{defect} c(T_G, E_{\text{FG}}, U_G)\, q(E_{\text{FG}}, U_G) = 0 \qquad (4)$$

Here, $n_h$ and $n_e$ are the density of free holes and electrons, respectively; $T_G$ is the growth temperature; $q(E_{\text{FG}}, U_G)$ is the defect charge state; $c(T_G, E_{\text{FG}}, U_G)$ is the defect concentration. It is worthy pointing out that solving Eqn. 4 at $T_G$ is critical to obtain the correct defect concentrations, because the high growth temperatures can dramatically impact $E_{FG}$.

When the external voltage is removed and temperature is reduced after growth or doping, the defect concentrations are expected to be largely frozen to the values under $T_G$ and $U_G$. However, the defect charge states, the concentrations of free charge carriers, and the Fermi level are supposed to re-equilibrate at the after-growth or after-doping temperature, $T_{AG}$. Therefore, the charge neutrality condition at $T_{AG}$ as expressed in Eqn. 5 should be executed.

$$n_h(T_{AG}, E_{\text{FAG}}) - n_e(T_{AG}, E_{\text{FAG}}) + \sum_{defect} c(T_G, E_{\text{FG}}, U_G)\, q(E_{\text{FAG}}) = 0, \qquad (5)$$

where $E_{\text{FAG}}$ and $q(E_{\text{FAG}})$ are the self-consistent Fermi level and defect charge state at the temperature of $T_{AG}$ without external voltage.

The above voltage-assisted-doping approach exhibits two favorable advantages. First, this approach is insensitive to the atomic composition of defects and growth conditions, because all the spontaneous donors or acceptors possess the same charge sign and thus undergo the same qualitative change by external voltage. Second, the external voltage can tune the defect formation energies without changing other growth conditions, such as growth temperature and precursors.

By contrast, the traditional methodology of tuning the defect formation energies by chemical potentials is largely limited by the optimal growth window.

### B. Analyses of Doping Asymmetry in ZnO

To verify the proposed approach, we choose ZnO as a touchstone, which is known for spontaneous heavy n-type doping [6,31,32], has abundant experiments for verification, and possesses wafer-size single crystals and versatile potential applications [33,34]. We carry out first-principles calculations of defect thermodynamics based on the density functional theory and closely examine three representative growth methods of ZnO. The first is the MBE growth method, which involves a growth temperature of ~690 K, possesses extremely low impurities in high vacuum, and very low oxygen partial pressure of ~$1.3\times10^{-7}$ atm [24]. The second is the hydrothermal growth method, which is carried out at ~650 K and involves ZnO solute in water under a vapor pressure of ~1085 atm [25]. In this case, water decomposition into hydrogen and oxygen should be considered. The third is the pressurized melt growth method, which involves melting and crystallization of ZnO at ~2200 K under a high oxygen partial pressure of ~50 atm [27]. It has been found that metal impurities, such as Pb, Cd and Fe, exist in the final crystals. See more details in the Methods.

Figure 3 shows the defect formation energies without external voltage for the relevant point and pair defects in ZnO under the MBE growth, hydrothermal growth, and pressurized melt growth conditions. Under the growth temperature of 692 K, ZnO crystal grown by MBE method is only lightly n-type-doped with a Fermi level 1.85 eV above the VBM, as shown in Fig. 3(a-b). The dominant donor is $Zn_i^{2+}$ with a concentration of $1.5\times10^{11}$ cm$^{-3}$, comparable to the free electron and hole concentrations of $2.9\times10^{11}$ and $1.7\times10^{10}$ cm$^{-3}$, respectively. When the temperature is reduced to the after-growth temperature of 300 K, the free holes are dramatically reduced to almost zero ($4.6\times10^{-30}$ cm$^{-3}$), but the concentration of $Zn_i^+$ and free electrons are largely frozen to their values at 692 K. This phenomenon finally leads to a significantly increased Fermi level of 2.94 eV, which corresponds to a strong n-type doping. Therefore, the n-type doping in the MBE-grown ZnO is not only caused by the Zn-rich growth condition (see Table 1 in the Methods), but also because the high growth temperature leads to a Fermi level around midgap, which intrinsically favors/disfavors the formation of donors/acceptors.

Figure 3(d-e) shows that the ZnO crystal prepared by hydrothermal growth method is also a strong n-type semiconductor, with an after-growth Fermi level at 3.01 eV above VBM, and thus confirms the experimental findings [26,35]. Different from the high-purity reactants in the MBE method,

the hydrogen impurities $H_i^+$ and $H_O^+$, with a respective concentration of $3.2\times10^{12}$ and $6.6\times10^7$ cm$^{-3}$, dominate the spontaneous n-type doping in the hydrothermal growth method. Other major defects include $(V_{Zn}+2H)^0$, $V_O^0$, and $H_{Zn}^-$. Because hydrogen is ubiquitous for the hydrothermal growth method, reduction of the hydrogen-related donors is quite challenging under the typical growth conditions.

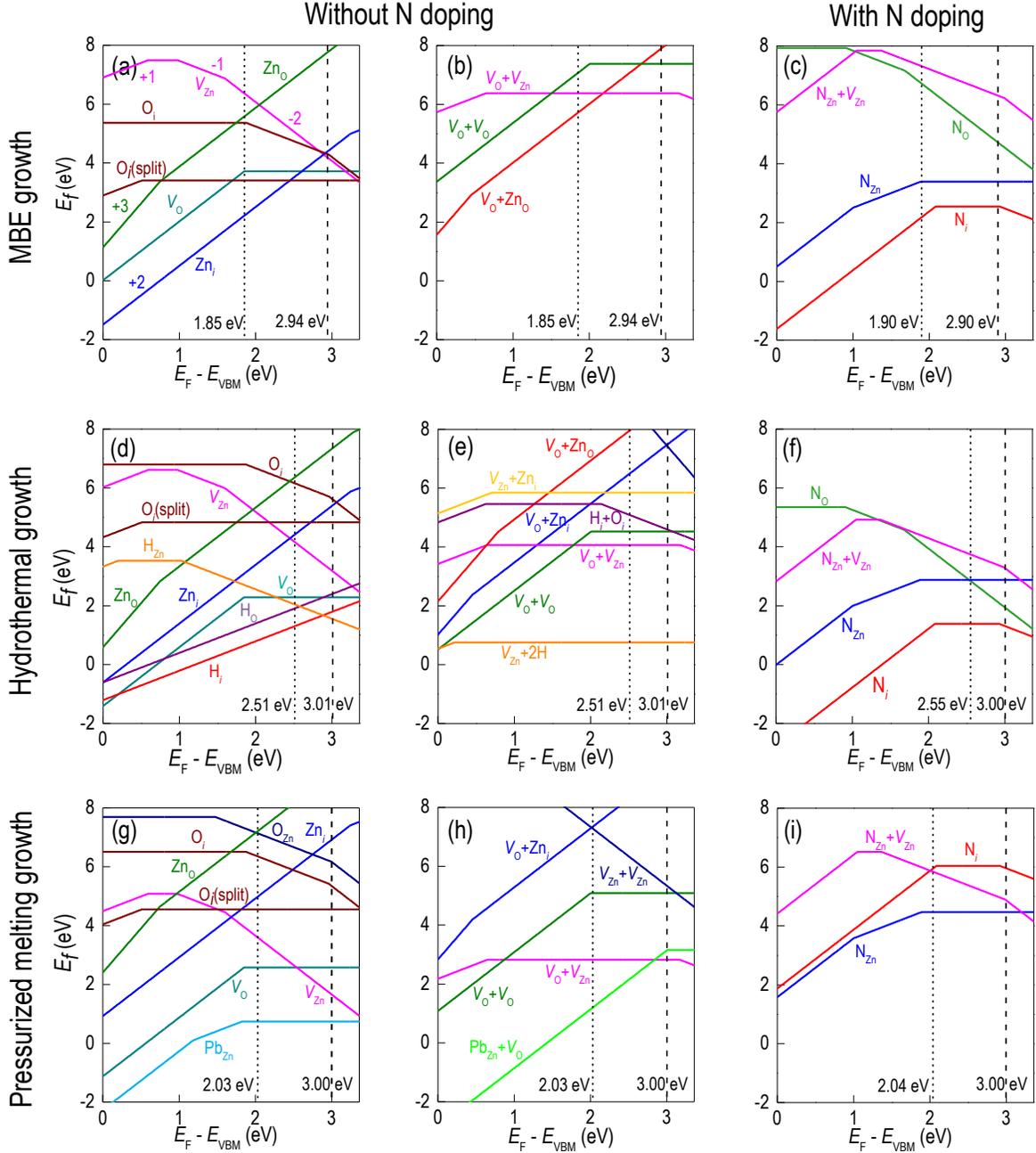

FIG. 3. Defect formation energy in ZnO for the (a-c) MBE growth, (d-f) hydrothermal growth, and (g-i) pressurized melt growth environment without external voltage. The point defects, pair defects, and nitrogen-dopant defects are shown in (a, d, g), (b, e, h), and (c, f, i), respectively, for easy visualization. The vertically dotted lines in (a, b, d, e, g, h) and (c, f, i) indicate self-consistent Fermi levels at growth temperatures without and with nitrogen doping, respectively, while the vertically dashed lines indicate those re-equilibrating at an after-growth temperature of 300 K.

Unlike the MBE and hydrothermal growth method, the pressurized melt growth involves a high oxygen partial pressure of ~50 atm and a high growth temperature of ~2200 K. Also, it has been reported that Pb and other impurities exist in the ZnO source material at a high level of ~5 ppm ($3.6 \times 10^{17}$ cm$^{-3}$) [27]. With consideration of the Pb-related defects, the Fermi level at the after-growth temperature is predicted to be 3.00 eV above VBM, as shown in Fig. 3(g-h), which explains the strong n-type doping effect in this growth method, too [36]. The major donor is found to be $(Pb_{Zn}+V_O)^{2+}$, with a concentration of $2.9 \times 10^{16}$ cm$^{-3}$. Other dominant defects include $Pb_{Zn}^0$, $V_O^0$, $(V_O+V_{Zn})^0$, and $V_{Zn}^{2-}$. If we simply take out the Pb impurity, the after-growth Fermi level changes to 2.03 eV, suggesting that reduction of Pb and other similar impurities will help mitigate the n-type doping in the pressurized melt growth method.

The above results clearly show that the dominant spontaneous donors vary significantly with the growth methods of ZnO. The native defect $Zn_i$ under the Zn-rich condition, the H-related defects from the water environment, and the metal impurities are responsible for the MBE method, the hydrothermal growth method, and the pressurized melt growth, respectively. We notice that H-related defects were suggested to be donors in ZnO from a general viewpoint without consideration of the growth conditions [37] and the other spontaneous donors mentioned above were totally missed, probably because of two reasons. First, the specific growth conditions were not considered in most previous predictions [38]. Second, the self-consistent determination of Fermi level with consideration of the growth and after-growth temperatures was not executed in literature.

Our calculations further indicate that p-type doping of ZnO without suppressing the abundant spontaneous donors are unlikely to succeed. Figure 3 (c, f, i) show the defect formation energies related to the nitrogen acceptor [31,39-42]. We find that, with the exactly same non-N defects, ZnO grown in these experiments still exhibits strong n-type doping, with a Fermi level of 2.90, 3.00, and 3.00 eV above the VBM for the MBE method, the hydrothermal growth method, and the pressurized melt growth method, respectively. The effect of nitrogen doping is almost negligible for all these growth methods and thus explain the unsuccessful p-type doping of ZnO with N [43]. Other p-type dopants, including P, As, and Ag, behave similarly as N [21].

## C. Application of Voltage-Assisted-Doping Approach to ZnO

We now employ the voltage-assisted-doping strategy to suppress the spontaneous donors and, meanwhile, enhance the acceptors in ZnO grown by the three types of experiment. Figure 4a shows that a $U_G$ of about -1.2 eV brings down the Fermi level from 2.94 to 1.85 eV above VBM for the MBE method. Meanwhile, the dominant donor $Zn_i$ is dramatically reduced from $1.5 \times 10^{11}$ to ~1 $cm^{-3}$, together with an acceptor increase of $O_i$ from ~$10^{-5}$ to ~0.1 $cm^{-3}$ and $V_{Zn}$ from $4.8 \times 10^{-11}$ to ~10 $cm^{-3}$ (Fig. 4b). Further increase of $U_G$ to about -1.5 eV can increase the acceptor $V_{Zn}$ to ~$10^6$ $cm^{-3}$ and reduce the Fermi level to 0.97 eV, a weak p-type doping region. Because $V_{Zn}$ possesses the lowest acceptor level at 0.97 eV among all the defects here (Fig. 3a), we reach the p-type doping limit for the MBE method. Note that the terraces of Fermi level around 1.85 and 0.97 eV (Fig. 4a) are caused by the neutral charge state of $O_i$ and $V_{Zn}$ below 1.85 and 0.97 eV (Fig. 3a), respectively.

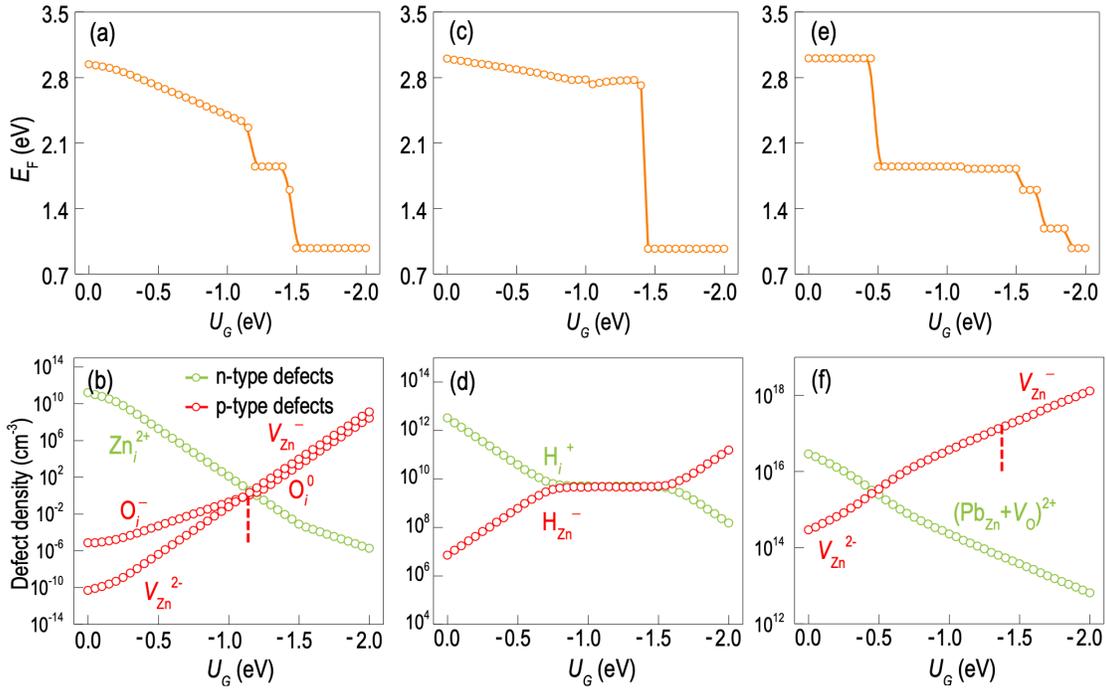

FIG. 4. Dependence of self-consistent Fermi level and dominant n- and p-type defects with the band bending $U_G$ for ZnO grown by the (a,b) MBE method, (c,d) hydrothermal growth method, and (e,f) pressurized melt growth method. The vertical red dashed lines in (b,f) indicate positions, where the defect charge states change. The after-growth temperature $T_{AG}$ is set to 300 K.

For the hydrothermal growth method, a $U_G$ of about -1.4 eV is capable to shift Fermi level to the p-type doping region at 0.97 eV (Fig. 4c). Meanwhile, the dominant donor $H_i^+$ is dramatically reduced from $3.2 \times 10^{12}$ to $4.5 \times 10^9$ cm$^{-3}$, and the dominant acceptor H$_{Zn}^-$ is increased from $7.1 \times 10^6$ to $4.8 \times 10^9$ cm$^{-3}$ (Fig. 4d). Because the lowest acceptor level of H$_{Zn}$ is at 1.04 eV (Fig. 3d), close to that of $V_{Zn}$, no further noticeable reduction of Fermi level can be achieved for this growth method. As for the pressurized melt growth method, a $U_G$ of about -0.5 eV decreases Fermi level to 1.84 eV (Fig. 4e), followed by a relatively flat region and then three jumps. The Fermi level limit here is same as those of the previous two methods at 0.97 eV. The dominant donor (Pb$_{Zn}$+$V_O$)$^{2+}$ is reduced from $2.8 \times 10^{16}$ to $6.4 \times 10^{12}$ cm$^{-3}$, and the dominant acceptor $V_{Zn}$ is increased from $2.8 \times 10^{14}$ to $1.2 \times 10^{18}$ cm$^{-3}$ (Fig. 4f). These results show that a proper positive band bending can efficiently decrease the spontaneous n-type defects in ZnO for all the three growth methods and meanwhile increase the p-type defects, leading to an ultimate limit of Fermi level at ~0.97 eV set by the lowest acceptor level of $V_{Zn}$.

To reveal the ultimate p-type doping limit of ZnO for the previously studied external dopants, we compare the defect energy levels (thermodynamic transition levels) of four p-type dopants for ZnO [43-46]: nitrogen, arsenic, phosphorus, and silver. Figure 5 shows that among all the thirteen defects, N$_O$ possesses the lowest acceptor level at 0.90 eV, which is slightly lower than that of $V_{Zn}$ at 0.97 eV. Therefore, none of these dopants can push the Fermi level of ZnO below 0.90 eV. To achieve a more efficient p-type doping, we envision the need to identify alternative dopants with even lower acceptor energy levels and employ the proposed voltage-assisted-doping method to enable an effective control over the defect concentrations.

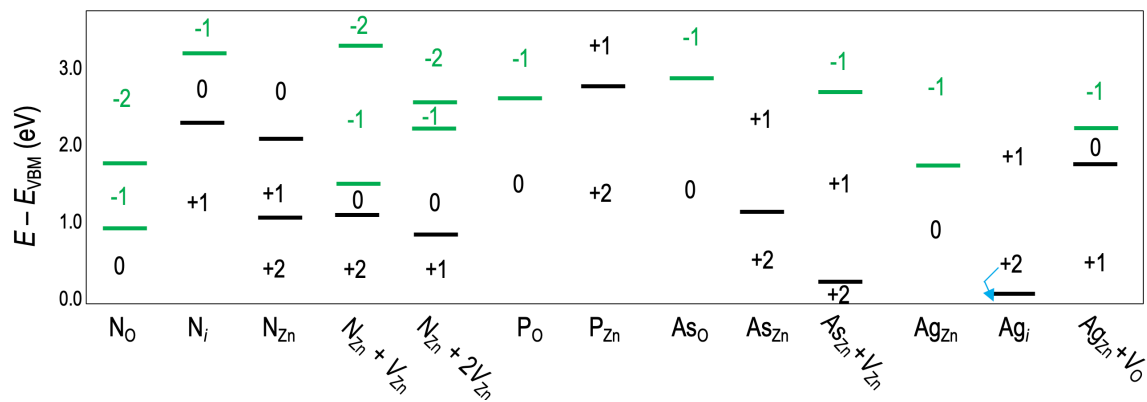

FIG. 5. Thermodynamic transition levels of nitrogen-, arsenic-, phosphorus-, and silver-related defects in ZnO.

### D. Experimental Realization of the Proposed Approach

We finally propose a schematic device to realize the voltage-assisted-doping method based on the edge-defined film-fed growth method [47]. As shown in Fig. 6a, an external voltage can be applied between the top and bottom electrodes in contact with crystal. For ZnO, a positive voltage is applied to the bottom electrode, where excess electrons accumulate and induce a downward band bending that reverses the doping asymmetry: the negatively charged or p-type defects are attracted/generated, while the positively charged or n-type defects are repelled/depressed. Although the band bending could be small at the beginning due to the initial n-type doping in the seed crystal, the external voltage can lead to a reduced n-type doping and subsequently a stronger band bending. When the growth continues and the crystal is pulled up, the band bending increases and the Fermi level decreases further (Fig. 6b). Because the growth front is kept around the bottom electrode, a homogeneous doping should be achieved, irrespective of the gradient of band bending in Fig. 2a. One final note is that the temperature gradient and growth speed should be controlled in a way that the defect diffusion is significantly lower than the growth speed times a critical length, namely, $D << v_z*l$, where $D$ is defect diffusivity, $v_z$ is the out-of-plane growth speed, and $l$ is the length of major band-bending region in which the growth front also locates, so the defects are largely frozen in the region of weak band bending.

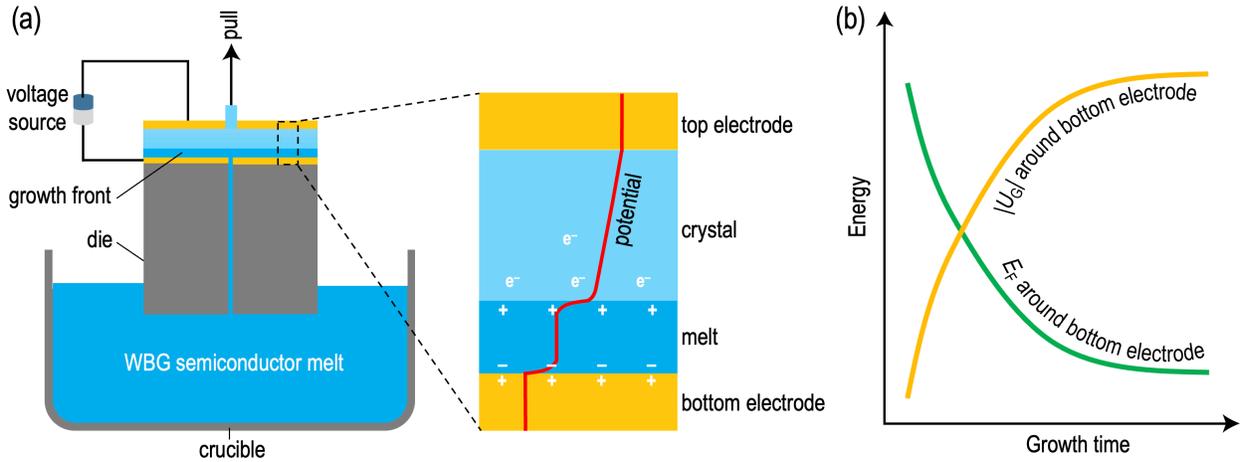

FIG. 6. (a) Proposed device to realize the voltage-assisted-doping approach and the schematic potential distribution across the electrodes; (b) schematic change of band bending and Fermi level with growth time for region near the positively charged bottom electrode in the proposed device.

## IV. Conclusion

In summary, we propose a voltage-assisted-doping approach to address the doping asymmetry in semiconductors by tunning the band edges and defect thermodynamics using external voltages. We demonstrate the capability of this method in ZnO based on density functional theory calculations. Our results show that the dominant spontaneous n-type defects in ZnO vary among the MBE method, the hydrothermal growth method, and the pressurized melt growth method, but the voltage-assisted-doping approach can dramatically reduce all these donors, increase the acceptors, and successfully brings the Fermi level to the p-type doping limits set by the lowest acceptor levels. Because the proposed approach is based on the fundamental defect thermodynamics, it should be applicable to the doping of semiconductors in general.

## Acknowledgments

This work was supported by the fund of the Guangdong Provincial Key Laboratory of Computational Science and Material Design (No. 2019B030301001), the Introduced Innovative R&D Team of Guangdong (2017ZT07C062), and the Shenzhen Science and Technology Innovation Committee (No. JCYJ20200109141412308). All the DFT calculations were carried out on the TaiYi cluster supported by the Center for Computational Science and Engineering of Southern University of Science and Technology and also on The Major Science and Technology Infrastructure Project of Material Genome Big-science Facilities Platform supported by Municipal Development and Reform Commission of Shenzhen.